\documentclass[aip,amsmath,amssymb,reprint]{revtex4-2}
\usepackage{graphicx}
\usepackage{dcolumn}
\usepackage{bm}
\usepackage[utf8]{inputenc}
\usepackage[T1]{fontenc}
\usepackage{mathptmx}
\usepackage[mathscr]{eucal}
\usepackage{etoolbox}
\usepackage{xcolor}

\makeatletter
\def\@email#1#2{%
 \endgroup
 \patchcmd{\titleblock@produce}
  {\frontmatter@RRAPformat}{\frontmatter@RRAPformat{\produce@RRAP{*#1\href{mailto:#2}{#2}}}\frontmatter@RRAPformat}
  {}{}
}%
\makeatother
\begin{document}


\title[]{Absence of cross-sublattice spin pumping and spin-transfer torques in collinear antiferromagnets}
\author{Junyu Tang}
\affiliation{Department of Physics and Astronomy, University of California, Riverside, California 92521, USA.}
\author{Ran Cheng}
\affiliation{Department of Electrical and Computer Engineering, University of California, Riverside, California 92521, USA.}
\affiliation{Department of Physics and Astronomy, University of California, Riverside, California 92521, USA.}

\begin{abstract}
We resolve the debate over the existence and magnitude of cross-sublattice (CS) contributions to spin pumping and spin-transfer torques in a two-sublattice antiferromagnet connected to a non-magnetic metal. Guided by symmetry considerations, we first relate the controversial CS terms to specific components in the spin conductance matrix. Then we quantify these components by studying the spin-dependent electron scattering on a fully compensated interface. We ascertain the absence of all CS contributions in the collinear regime. Even in the non-collinear regime, the CS contributions only constitute a higher-order correction to the existing theory.
\end{abstract}

\maketitle

\section{Introduction}
Demystifying the intricate interplay between magnetic excitations and electronic transport is essential for realizing efficient electric control of magnetism, which underlies recent development of spintronics, especially the emerging frontier of sub-terahertz spintronics exploiting the unique dynamics of antiferromagnets (AFMs)~\cite{RMPAFM,Han2023coherent}. From a fundamental  perspective, excitations of magnetic order can be converted into pure spin currents of electrons either coherently or incoherently. While an incoherent spin generation involves an ensemble of thermal magnons to exchange spin angular momenta with electrons, a coherent spin generation, on the other hand, typically involves only a resonance mode and is achieved by virtue of spin pumping~\cite{SpinPumping1,SpinPumping2}.

It has been shown that when a collinear AFM characterized by two unit sublattice-magnetic vectors $\bm{m}_A$ and $\bm{m}_B$ is interfaced with a non-magnetic metal (NM), the coherent dynamics of the Néel vector $\bm{n}=(\bm{m}_A-\bm{m}_B)/2$ and of the small magnetic moment $\bm{m}=(\bm{m}_A+\bm{m}_B)/2$ can pump a total pure spin current into the NM in the form of~\cite{Cheng2014PRL,Johansen2017PRB,GomonayPhenom2010PRB}
\begin{align}
    \frac{e}{\hbar}\bm{I}_s = G^r (\bm{n}\times \dot{\bm{n}}+\bm{m}\times \dot{\bm{m}})-G^i \dot{\bm{m}}, \label{eq:Is_col_AFM_mn}
\end{align}
where $\bm{I}_s$ is measured in units of an electric current (in \textit{Amp}), $e$ is the absolute electron charge, $\hbar$ is the reduced Planck constant, $G^r$ and $G^i$ are two independent components of
the interfacial spin conductance~\cite{clarifyspinmixing} which can be rigorously calculated by considering the microscopic spin-dependent scattering on the AFM/NM interface~\cite{Cheng2014PRL,Cheng2014aspects}. Equation~\eqref{eq:Is_col_AFM_mn} can be equivalently written in terms of the sublattice-magnetic vectors as
\begin{align}
    \frac{e}{\hbar}\bm{I}_s = G^{AA} \bm{m}_A\times \dot{\bm{m}}_A +G^{BB} \bm{m}_B\times \dot{\bm{m}}_B -G^A \dot{\bm{m}}_A -G^B\dot{\bm{m}}_B, \label{eq:Is_col_AFM_AB}
\end{align}
where $G^{AA}=G^{BB}=G^r/2$ and $G^A=G^B=G^i/2$ are the spin conductance components associated with each sublattice. 

Recently, coherent spin pumping in collinear AFMs that are described by the above equations has been experimentally verified in a number of materials, notably in MnF$_2$~\cite{vaidya2020subterahertz}, Cr$_2$O$_3$~\cite{Li2020spin}, $\alpha-$Fe$_2$O$_3$~\cite{Wang2021PRL,Lebrun2021PRL}, and synthetic AFM~\cite{SyntheticAFM2022NC}, stimulating a vibrant search of new physics in the sub-terahertz frequency range harnessing the unique spin dynamics of AFMs.

Latest theoretical studies~\cite{KamraAFM,KamraTwoSublattice,CrossSublattice,Yuan2019proper}, however, suggest that Eq.~\eqref{eq:Is_col_AFM_AB} should also admit cross-sublattice (CS) terms $G^{AB}\bm{m}_A\times \dot{\bm{m}}_B$ and $G^{BA}\bm{m}_B\times \dot{\bm{m}}_A$, which in turns changes the Gilbert damping constant into a matrix.
While such terms do not contradict directly with existing experimental observations, they could modify the strength of spin pumping predicted by Eqs.~\eqref{eq:Is_col_AFM_mn} and~\eqref{eq:Is_col_AFM_AB}, thus affecting the numerical extraction of the interfacial spin conductance involved in different materials. Even more surprising is that the CS terms are believed to exist even for a fully compensated AFM/NM interface in the collinear limit~\cite{KamraAFM}. Furthermore, in the non-collinear regime of a two-sublattice AFM (e.g., the spin-flop phase induced by a strong magnetic field), the established form of spin pumping is questionable, especially whether $\bm{n}\times\dot{\bm{n}}$ and $\bm{m}\times\dot{\bm{m}}$ in Eq.~\eqref{eq:Is_col_AFM_mn} still share the same coefficient~\cite{ReitzTserkovnyak2020,RodriguezBarsukov}. In direct connection with the CS spin pumping, CS spin-transfer torques are allowed by the Onsager reciprocal relations~\cite{KamraTwoSublattice,CrossSublattice,Yuan2019proper}, but their existence remains experimentally elusive.

In this Letter, we resolve the puzzle of CS contributions to spin pumping and spin-transfer torques in collinear AFMs from a theoretical perspective. Guided by phenomenological argument, we first clarify a number of essential mathematical relations between spin pumping and spin-transfer torques in the presence of CS contributions, where the controversial CS components are separated and expressed in terms of the corresponding coefficients in the spin conductance matrix. Then we calculate these coefficients by studying the microscopic spin-dependent scattering of electrons off a fully compensated AFM/NM interface. We claim that all CS effects vanish in the collinear regime for fully compensated interfaces, affirming the validity of the established theories [\textit{viz.} Eqs.~\eqref{eq:Is_col_AFM_mn} and~\eqref{eq:Is_col_AFM_AB}] and the experimental fitting they enable. We find that even in the non-collinear regime, the CS effects only bring about higher-order corrections.

\section{Phenomenological relations}
In its most general form, the coherent spin pumping by a two-sublattice AFM into an adjacent NM can be written as
\begin{align}
    \frac{e}{\hbar}\bm{I}_s=
    (\bm{m},\bm{n})\times
    \begin{pmatrix}
    G^{mm} & G^{mn}\\
    G^{nm} & G^{nn}
    \end{pmatrix}
    \begin{pmatrix}
    \dot{\bm{m}}\\
    \dot{\bm{n}}
    \end{pmatrix}-G^m\dot{\bm{m}},
    \label{eq:AFM_pump_general}    
\end{align}
which differs (is generalized) from Eq.~\eqref{eq:Is_col_AFM_mn} by the off-diagonal terms $G^{mn}$ and $G^{nm}$ in the matrix of interfacial spin conductance. The dynamics of the AFM can be described by a set of coupled Landau–Lifshitz equations as~\cite{AFMLLG}
\begin{subequations} 
\label{eq:phenom_LLG}
\begin{align}   
 \hbar \dot{\bm{m}} &= \bm{f}^n\times \bm{n}+ \bm{f}^m\times \bm{m}, \\
 \hbar \dot{\bm{n}} &= \bm{f}^m\times \bm{n}+ \bm{f}^n\times \bm{m},
\end{align}
\end{subequations}
where the Gilbert damping is omitted for simplicity, $\bm{f}^m = -\partial\epsilon/\partial\bm{m}$ and $\bm{f}^n = -\partial\epsilon/\partial\bm{n}$ are the effective fields (or driving forces) with $\epsilon$ being the magnetic free energy. In our convention, $\bm{f}^m$ and $\bm{f}^n$ are scaled in units of energy. By inserting Eqs.~\eqref{eq:phenom_LLG} into Eq.~\eqref{eq:AFM_pump_general}, we can relate $\bm{I}_s$ to these driving forces and establish a linear response relation
\begin{align}
 I_{s,i}=L^{sm}_{ij} f^m_j+L^{sn}_{ij}f^n_j,\quad (i,j \mbox{ run over } x,y,z)
\end{align}
where the response coefficients $L^{sm}_{ij}(\bm{m},\bm{n})$ and $L^{sn}_{ij}(\bm{m},\bm{n})$ are related to the spin conductance. As the inverse effect of spin pumping, the spin-transfer torques can be expressed as $T^m_i=L^{ms}_{ij}V_j^s$ and $T^n_i=L^{ns}_{ij}V_j^s$, where $\bm{V}^s=\bm{\mu}_s/e$ is the spin voltage with $\bm{\mu}_s=(\mu_{\uparrow}-\mu_{\downarrow})\hat{\bm{s}}$ being the spin chemical potential ($\hat{\bm{s}}$ specifies the quantization axis). These response coefficients must satisfy the Onsager reciprocal relation
\begin{align}
 L^{ms}_{ij}(\bm{m},\bm{n})=L^{sm}_{ji}(-\bm{m},-\bm{n}),
 \label{eq:Onsager}
\end{align}
as both $\bm{m}$ and $\bm{n}$ break the time-reversal symmetry. An identical relation is applicable to $L^{ns}_{ij}$ and $L^{sn}_{ji}$ as well. When $\bm{V}_s$ is treated as a common driving force, $L^{ms(ns)}$ and $L^{sm(sn)}$ will share the same unit, which simplifies the following discussions. After some straightforward algebra, we find
\begin{subequations}
\label{eq:TmTn}
\begin{align}
e\bm{T}^m=&G^{mm}\bm{m}\times(\bm{V}_s\times\bm{m})+G^{mn}\bm{n}\times(\bm{V}_s\times\bm{m}) \notag\\
&+G^{nm}\bm{m}\times(\bm{V}_s\times\bm{n})+G^{nn}\bm{n}\times(\bm{V}_s\times\bm{n}) \notag\\
&+G^m\bm{m}\times\bm{V}_s, \\  e\bm{T}^n=&G^{mm}\bm{n}\times(\bm{V}_s\times\bm{m})+G^{mn}\bm{m}\times(\bm{V}_s\times\bm{m}) \notag\\
&+G^{nm}\bm{n}\times(\bm{V}_s\times\bm{n})+G^{nn}\bm{m}\times(\bm{V}_s\times\bm{n}) \notag\\
&+G^m\bm{n}\times\bm{V}_s,
\end{align}
\end{subequations}
where all spin-transfer torques have been scaled into the inverse-time dimension so that $\bm{T}^{m}$ and $\bm{T}^{n}$ can be directly added to Eqs.~\eqref{eq:phenom_LLG}.

The spin-transfer torques exerting on the two sublattice-magnetic moments, $\bm{m}_A$ and $\bm{m}_B$, are $\bm{T}^A=\bm{T}^m+\bm{T}^n$ and $\bm{T}^B=\bm{T}^m-\bm{T}^n$. A simple manipulation of Eq.~\eqref{eq:TmTn} shows that
\begin{subequations}
\label{eq:torque}
\begin{align}
    \bm{T}^A=&\tau^{AA}_{D} \bm{m}_A\times (\bm{V}^s\times \bm{m}_A)+\tau^{AB}_{CS} \bm{m}_A\times (\bm{V}^s\times \bm{m}_B)\nonumber\\ 
    &+\tau^A_F \bm{m}_A\times \bm{V}^s,\\
    \bm{T}^B=&\tau^{BB}_{D} \bm{m}_B\times (\bm{V}^s\times \bm{m}_B)+\tau^{BA}_{CS} \bm{m}_B\times (\bm{V}^s\times \bm{m}_A)\nonumber\\
    &+\tau^B_F \bm{m}_B\times \bm{V}^s,
\end{align}
\end{subequations}
where $\tau_{D}^{AA(BB)}$, $\tau^{AB(BA)}_{CS}$ and $\tau_F^{A(B)}$ represent the coefficients of the damping-like torques, the CS torques, and the field-like torques, respectively. To relate these torques to the spin conductance matrix appearing in Eq.~\eqref{eq:AFM_pump_general}, we now define $G^{AA(BB)}=e\tau^{AA(BB)}_D/2$, $G^{AB(BA)}=e\tau^{AB(BA)}_{CS}/2$, and $G^{A(B)}=e\tau^{A(B)}_F$, which satisfy
\begin{subequations}
\label{eq:torque_coeff}
    \begin{align}
    G^{AA} &= \frac{1}{4}(G^{mm}+G^{mn}+G^{nm}+G^{nn}), \\
    G^{BB} &= \frac{1}{4}(G^{mm}-G^{mn}-G^{nm}+G^{nn}), \\
    G^{AB} &= \frac{1}{4}(G^{mm}+G^{mn}-G^{nm}-G^{nn}), \\
    G^{BA} &= \frac{1}{4}(G^{mm}-G^{mn}+G^{nm}-G^{nn}),
    \end{align}
\end{subequations}
and $G^A = G^{B} = G^m$. By invoking the Onsager reciprocal relations, we obtain
\begin{align}
    \frac{e}{\hbar}\bm{I}_s = 
    (\bm{m}_A,\bm{m}_B)&\times
    \begin{pmatrix}
    G^{AA} & G^{AB}\\
    G^{BA} & G^{BB}
    \end{pmatrix}
    \begin{pmatrix}
    \dot{\bm{m}}_A\\ \dot{\bm{m}}_B
    \end{pmatrix} -G^A\dot{\bm{m}}_A-G^B\dot{\bm{m}}_B,
    \label{eq:AFM_pump_sublattice}    
\end{align}
which incorporates Eq.~\eqref{eq:Is_col_AFM_AB} as a special case when the CS terms vanish ($G^{AB}=G^{BA}=0$). Because combining Eqs.~\eqref{eq:torque_coeff} and~\eqref{eq:AFM_pump_sublattice} can reproduce Eq.~\eqref{eq:AFM_pump_general} under the definitions of $\bm{n}=(\bm{m}_A-\bm{m}_B)/2$ and $\bm{m}=(\bm{m}_A+\bm{m}_B)/2$, we have established consistent relations between spin pumping and spin-transfer torques in the presence of CS contributions, which hold in both the $(\bm{m},\bm{n})$ basis and the $(\bm{m}_A,\bm{m}_B)$ basis.

By imposing symmetry constraints on $\bm{m}_A$ and $\bm{m}_B$ (or $\bm{m}$ and $\bm{n}$), one can reduce the number of independent variables in the matrix of spin conductance. For an insulating AFM, practically only the magnetic layer in direct contact with the NM is relevant. For an uncompensated interface such as the $(111)$ plane of NiO, the conduction electrons only couple to one sublattice so the CS terms become irrelevant. Spin pumping in this special case is physically equivalent to its ferromagnetic counterpart~\cite{Cheng2014PRL,Cheng2014aspects}. On the contrary, for a fully compensated interface, there is an effective $PT$ symmetry requiring that $\tau^{AA}_{D}=\tau^{BB}_{D}$ and $\tau_{CS}^{AB}=\tau_{CS}^{BA}$, which, according to Eqs.~\eqref{eq:torque_coeff}, yields $G^{mn}=G^{nm}=0$. What deserves special consideration is the case of partially compensated interfaces, which typically arises when a compensated surface is subject to interfacial roughness, nonequivalent electron couplings to the two sublattices, and so on, breaking the effective $PT$ symmetry. In the presence of $\bm{T}^A$ and $\bm{T}^B$ defined by Eqs.~\eqref{eq:torque}, the effective dynamics of the AFM can be formally written as
\begin{subequations}
 \begin{align}
  \hbar\dot{\bm{m}}_A=(\bm{f}_{AA}+\bm{f}_{AB})\times\bm{m}_A, \\
  \hbar\dot{\bm{m}}_B=(\bm{f}_{BA}+\bm{f}_{BB})\times\bm{m}_B,
 \end{align}
\end{subequations}
where the diagonal fields $\bm{f}_{AA(BB)}$ involve the single-ion anisotropy, the external Zeeman field and the ordinary spin torques $\tau_D^{AA(BB)}$ and $\tau_F^{A(B)}$, whereas the off-diagonal fields $\bm{f}_{AB(BA)}$ include the exchange interaction between $\bm{m}_A$ and $\bm{m}_B$ as well as the CS torques $\tau_{CS}^{AB}$ and $\tau_{CS}^{BA}$. Regarding $\bm{V}^s\times\bm{m}_B$ and $\bm{V}^s\times\bm{m}_A$ as two reciprocal driving forces cross-linking $\bm{m}_A$ with $\bm{m}_B$, we could apply the Onsager reciprocity relation to $\bm{f}_{AB}$ and $\bm{f}_{BA}$, which gives rise to $\tau_{CS}^{AB}=\tau_{CS}^{BA}$ even though $\tau_{CS}^{AA}$ could differ from $\tau_{CS}^{BB}$. This means that on a partially compensated interface without the effective $PT$ symmetry, we still have $G^{mn}=G^{nm}$ (hence $G^{AB}=G^{BA}$) based on the relations in Eqs.~\eqref{eq:torque_coeff}, but this time $G^{mn}$ could be finite (hence $G^{AA}$ and $G^{BB}$ are different). Consequently, for compensated AFM/NM interfaces, we have the general relations 
\begin{align}
 &\tau^{AA (BB)}_{D}=\frac{1}{2e}(G^{mm}+G^{nn}\pm 2G^{mn}), \label{eq:ABfinal} \\
 &\tau^{AB}_{CS}=\tau^{BA}_{CS}=\frac{1}{2e}(G^{mm}-G^{nn}), \label{eq:CSfinal}
\end{align}
where $G^{mn}\rightarrow0$ and $\tau^{AA}_{D}\rightarrow\tau^{BB}_{D}$ as the interface becomes fully compensating. In any case, the CS terms could only exist when $G^{mm}\neq G^{nn}$. Therefore, to resolve the puzzle of CS contributions, we need to quantify and compare $G^{mm}$ and $G^{nn}$. Note, however, that if the two sublattices are chemically nonequivalent, the above analysis is invalid, which belongs to the category of ferrimagnets~\cite{GuoFiM}.

\section{Microscopic calculations}
\begin{figure}[t]
\includegraphics[width=\linewidth]{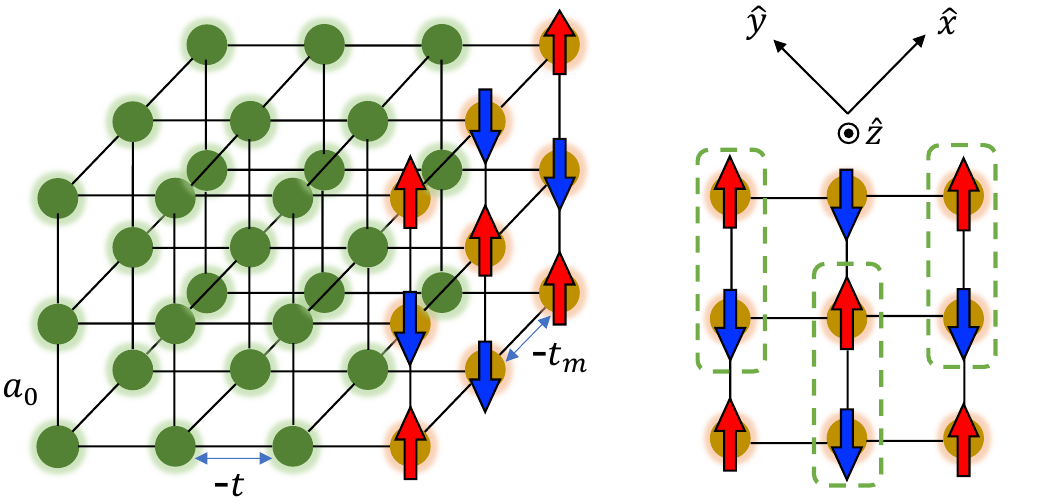}
\caption{A fully compensated AFM/NM interface with cubic lattice, where $t_m$ ($t$) is the hopping energy in the AFM (NM), and $a_0$ is the lattice constant. On the interface plane, the magnetic unit cell is indicated by green dashed circles, which are periodic in both the $\hat{x}$ ([1,1,0]) and $y$ ([1,-1,0]) directions.}
\label{fig:interface}
\end{figure}

The microscopic origin of $G^{mm}$ and $G^{nn}$ pertains to the spin-dependent scattering of electrons off an AFM/NM interface. Without losing generality, we consider a fully compensated interfacce with a simple cubic lattice~\cite{notelattice} as schematically illustrated in Fig.~\ref{fig:interface}. Here we ignore interfacial roughness because it incurs a random spin distribution that destroys the periodic magnetic structure in the lateral dimension, thus invalidating any meaningful band calculation. Following the wavefunction matching approach detailed in Chapter 4 of Ref.~\cite{Cheng2014aspects}, we shall determine the scattering matrix $\mathbb{S}$ of the form
\begin{align}
\label{eq:scatter}
    \mathbb{S}=\begin{pmatrix}
        \mathbb{S}^{++}& \mathbb{S}^{+-}\\
        \mathbb{S}^{-+} & \mathbb{S}^{--}
    \end{pmatrix},
\end{align}
where each block is a $2\times2$ matrix in the spin space and $\pm$ accounts for the sublattice (pseudo-spin) degree of freedom. Under the adiabatic condition (\textit{i.e.}, the dynamics of $\bm{m}$ and $\bm{n}$ is much slower than the electron relaxation), we have~\cite{Notelinear}
\begin{subequations}
\label{eq:scatter_decompose}
\begin{align}
    \mathbb{S}^{\pm\pm} \approx S&_0^{\pm\pm}\sigma_0+S_m^{\pm\pm}\bm{m}\cdot \bm{\sigma},\\
    \mathbb{S}^{\pm\mp} \approx S&_n\bm{n}\cdot \bm{\sigma}\pm S_{mn}(\bm{m}\times \bm{n})\cdot\bm{\sigma},
\end{align}
\end{subequations}
where $\bm{\sigma}=\{\sigma_x,\sigma_y,\sigma_z\}$ is the vector of Pauli spin matrices and $\sigma_0$ is the identity matrix. The complete expressions of $S_0^{\pm\pm}$, $S_m^{\pm\mp}$, $S_n$ and $S_{mn}$ in terms of the crystal momentum and other material parameters are provided in the Supplementary Material~\cite{SM}. If we turn to the collinear regime, $|\bm{n}^2|\approx1$ and $|\bm{m}^2|\ll1$, only $S_0^{\pm\pm}$ and $S_n$ in Eqs.~\eqref{eq:scatter_decompose} will remain essential, then the spin-flip scattering will be necessarily accompanied by the reversal of pseudo-spin~\cite{HaneyMacD}. If we go beyond the collinear regime, however, the locking between spin and pseudo-spin will be lifted. We also notice that previous studies assumed $S_m^{\pm\mp}\approx S_n$ in the collinear regime without a rigorous justification~\cite{Cheng2014PRL,Cheng2014aspects}, so here in a general context we treat all components in Eq.~\eqref{eq:scatter_decompose} as independent quantities.

The pumped spin current polarized in the $j$ direction ($j=x,y,z$) can be calculated by~\cite{Cheng2014PRL,Cheng2014aspects}
\begin{align}
 I_{s,j}=-\frac{e}{2\pi} {\rm{Im}}\left\{ {\rm{Tr}}\left[\mathbb{S}^\dagger(\sigma_0\otimes \sigma_j)\dot{\mathbb{S}}\right]\right\},
\end{align}
which, after some tedious algebra, ends up with
\begin{align}
    \frac{e}{\hbar} \bm{I}_s=G^{nn}_0\bm{n}\times\dot{\bm{n}}+G^{mm}_0\bm{m}\times\dot{\bm{m}}-G^m \dot{\bm{m}}+\Delta \bm{I}_s,
    \label{eq:Is_and_correction}
\end{align}
where
\begin{align}
    \Delta\bm{I}_s&=\Delta G\{ [\bm{n}\cdot (\bm{m}\times \dot{\bm{m}})]\bm{n}+\bm{m}\cdot (\bm{n}\times \dot{\bm{n}})]\bm{m}\} \notag\\
    &=\Delta G\left[(\bm{n}\times\dot{\bm{n}})|\bm{m}|^2+(\bm{m}\times\dot{\bm{m}})|\bm{n}|^2\right. \notag\\
    &\qquad\qquad\left.+2(\bm{n}\times\bm{m})(\bm{m}\cdot\dot{\bm{n}})\right]
\end{align}
is a (higher-order) correction term never claimed before. After a straightforward re-ordering of terms, Eq.~\eqref{eq:Is_and_correction} can be effectively written as 
\begin{align}
    \frac{e}{\hbar} \bm{I}_s=&(G^{nn}_0+|\bm{m}|^2\Delta G)\bm{n}\times\dot{\bm{n}}+(G^{mm}_0+|\bm{n}|^2\Delta G)\bm{m}\times\dot{\bm{m}} \notag\\
    &\quad\qquad-G^m \dot{\bm{m}}+2\Delta G (\bm{n}\times\bm{m})(\bm{m}\cdot \dot{\bm{n}}),
    \label{eq:Is_after_correction}
\end{align}
where $G^{mn}$ and $G^{nm}$ indeed vanish as required by the $PT$ symmetry; they could appear only as a consequence of imperfections (such as roughness) on a compensated interface. Eq.~\eqref{eq:Is_after_correction} is a generalized spin pumping formula involving four independent components of the interfacial spin conductance. These terms are determined by the corresponding components of the scattering matrix as
\begin{subequations}
\label{eq:G}
    \begin{align}
        &G^{nn}_0=\frac{e^2\mathscr{A}}{h\pi^2}\int |S_n|^2\mathrm{d}^2\bm{k},\\
        &G^{mm}_0=\frac{e^2\mathscr{A}}{2h\pi^2}\int\left(|S_m^{++}|^2+|S_m^{--}|^2\right)\mathrm{d}^2\bm{k},\\
        &G^{m}=\frac{e^2\mathscr{A}}{2h\pi^2}\int {\rm Im}\left[(S_0^{++})^*S_m^{++}+(S_0^{--})^*S_m^{--}\right]\mathrm{d}^2\bm{k},\\
        &\Delta G =\frac{e^2\mathscr{A}}{h\pi^2}\int |S_{mn}|^2\mathrm{d}^2\bm{k},        
    \end{align}
\end{subequations}
where $\mathrm{d}^2\bm{k}=dk_xdk_y$ and $\mathscr{A}$ is the area of the interface. Except the last term proportional to $\Delta G$ in Eq.~\eqref{eq:Is_after_correction}, we can read off the effective spin conductance $G^{nn}$ and $G^{mm}$ as
\begin{subequations}
 \begin{align}
  G^{nn}=G^{nn}_0+|\bm{m}|^2\Delta G, \\
  G^{mm}=G^{mm}_0+|\bm{n}|^2\Delta G,
 \end{align}
\end{subequations}
and, according to Eqs.~\eqref{eq:ABfinal} and~\eqref{eq:CSfinal}, we obtain the damping-like torques and the CS torques as
\begin{subequations}
\label{eq:tau_corrected}
\begin{align}    
\tau^{AA}_{D}=\tau^{BB}_{D}&=\frac{1}{2e}(G^{mm}_0+G^{nn}_0+\Delta G), \label{eq:tauDL_corrected}\\
\tau^{AB}_{CS}=\tau^{BA}_{CS}&=\frac{1}{2e}\left[G^{mm}_0-G^{nn}_0+(|\bm{n}|^2-|\bm{m}|^2)\Delta G\right], \label{eq:tauCS_corrected}
\end{align}
\end{subequations}
where $|\bm{n}|^{2}+|\bm{m}|^2=1$ is used. The above results are valid even in the noncollinear regime.

Finally, we point out that the last term in Eq.~\eqref{eq:Is_after_correction}, pursuant to the Onsager reciprocal relations, gives rise to an additional CS torque
\begin{align}
    \Delta\bm{T}^A=\Delta\bm{T}^B=\frac{\Delta G}{e}[\bm{V}^s\cdot (\bm{m}_1\times \bm{m}_2)](\bm{m}_2\times\bm{m}_1), \label{seq:delta_tau}
\end{align}
which is nonlinear in $\bm{m}_1\times\bm{m}_2$ thus not being captured by the phenomenological consideration in the previous section. In the collinear regime, this term is negligible.

\section{Numerical results}

\begin{figure}[t]
\includegraphics[width=0.9\linewidth]{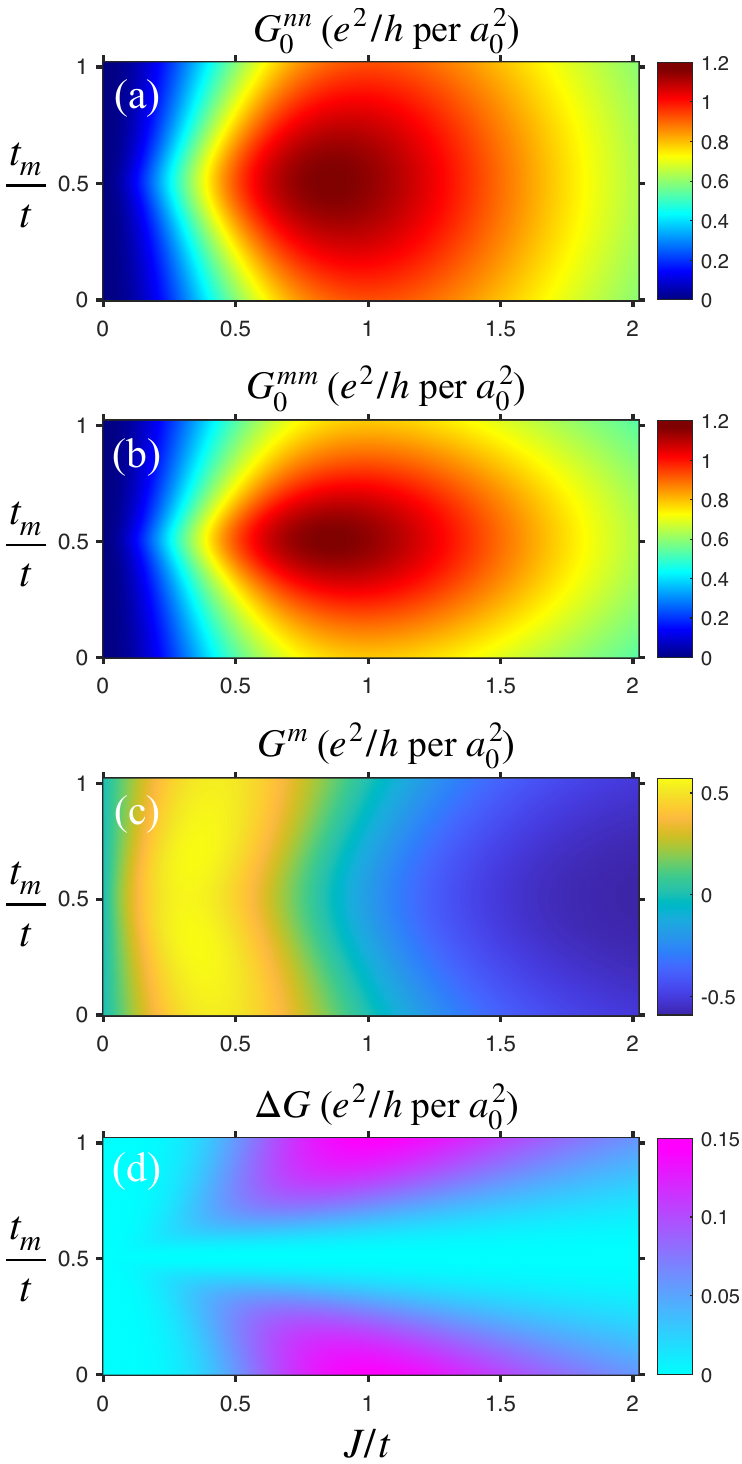}
\caption{Interfacial spin conductance per unit-cell area ($e^2/h$ per $a_0^2$) as a function of $J/t$ and $t_m/t$. Here, $J$ is the exchange coupling between the conduction electrons and the magnetic moments on the compensated interface, and $t_m$ ($t$) is the hopping energy on the AFM (NM) side of the interface.}
\label{fig:spin-mixing}
\end{figure}

In Fig.~\ref{fig:spin-mixing}, we numerically plot the four relevant components of the spin conductance basing on Eq.~\eqref{eq:G} as functions of the exchange coupling $J$ (between the conduction electrons and the magnetic moments) and the ratio of kinetic energies in the AFM and NM (\textit{i.e.}, hopping integrals $t_m$ and $t$). Here, the spin conductance is expressed in units of $e^2/h$ per $a_0^2$ (area of a magnetic unit cell on the interface), which should be multiplied by the number of magnetic unit cells $\mathscr{N}$ on the interface to retrieve the total spin conductance. Comparing Fig.~\ref{fig:spin-mixing}(a) and (b), we find that $G^{nn}_0$ and $G^{mm}_0$ share a very similar pattern as they both culminate around $J/t=1$ and $t_m/t=0.5$. They are the dominant contributions to the damping-like torques [see Eq.~\eqref{eq:tauDL_corrected}]. Figure~\ref{fig:spin-mixing}(c) for $G^m$, on the other hand, shows how the strength of field-like torques varies over $J$ and $t_m$. It is clear that the damping-like (field-like) torques dominate the strong (weak) exchange coupling regime, which is corroborated by a recent experiment~\cite{KentExp}.

In the collinear limit that $|\bm{n}|^2\rightarrow 1$ and $|\bm{m}|^2\rightarrow0$, the CS torques, according to Eq.~\eqref{eq:tauCS_corrected}, reduce to
\begin{align}
 \tau^{AB(BA)}_{CS}=\frac1{2e}(G^{mm}_0+\Delta G-G^{nn}_0).
\end{align}
Using the numerical results shown in Fig.~\ref{fig:spin-mixing}, we find that
\begin{align}
    G^{nn}_0=G^{mm}_0+\Delta G,
    \label{eq:equality}
\end{align}
which renders all CS torques exactly zero. As a matter of fact, regarding the integrands in Eqs.~\eqref{eq:G} as functions of the crystal momentum, one can rigorously prove that Eq.~\eqref{eq:equality} is an exact identify (see details in the Supplementary Material~\cite{SM}). It is interesting that even in the highly non-collinear regime where $|\bm{m}^2|$ is comparable to $|n|^2$, the CS torques [proportional to $G^{mm}_0-G^{nn}_0+(|\bm{n}|^2-|\bm{m}|^2)\Delta G$] is at most a few percents of the damping-like torques since $\Delta G$ is much smaller than $G_0^{nn}$ and $G_0^{mm}$, as shown in Fig.~\ref{fig:spin-mixing}(d).

In conclusion, we have justified the absence of CS contributions to the coherent spin pumping and spin-transfer torques in the collinear regime of two-sublattice AFMs, substantiating the data extraction and fitting approach based on Eqs.~\eqref{eq:Is_col_AFM_mn} and~\eqref{eq:Is_col_AFM_AB} in explaining a number of recent experiments~\cite{vaidya2020subterahertz,Li2020spin,Wang2021PRL,Lebrun2021PRL,SyntheticAFM2022NC}.

\begin{acknowledgments}
The authors acknowledge fruitful discussions with Hantao Zhang. This work is support by the Air Force Office of Scientific Research (Grant No. FA9550-19-1-0307). 
\end{acknowledgments}

\section*{Supplementary Material}
See the supplementary materials for more mathematical details about the interfacial spin conductance.

\section*{Data Availability Statement}
The data that support the findings of this study are available from the corresponding author upon reasonable request.

\nocite{*}
\bibliography{Reference}

\begin{thebibliography}{27}
\bibitem [{cla()}]{clarifyspinmixing}%
  \BibitemOpen
  \href@noop {} {}\bibinfo {note} {{In literature, $G^r$ and $G^i$ are often
  dubbed the real and imaginary parts of the ``spin-mixing'' conductance, which
  we believe is not an accurate nomenclature. In an AFM, the spin and
  sublattice degrees of freedom are entangled, so ``spin-mixing'' does not
  fully capture the underlying physics, if not entirely misleading. From the
  perspective of spin-dependent scattering, $G^r$ and $G^i$ are related to
  different components of the scattering matrix, which cannot be clearly
  distinguished by ``real'' and ``imaginary'' (even though they are true in
  ferromagnets). Therefore, we elect to just call them different components of
  the (interfacial) spin conductance.}}\BibitemShut {Stop}%
\bibitem [{Not()}]{Notelinear}%
  \BibitemOpen
  \href@noop {} {}\bibinfo {note} {Here we have truncated the iterative
  solution at the linear order in $\bm{m}$, which has a magnitude $|\bm{m}|<1$
  even in the non-collinear regime.}\BibitemShut {Stop}%
\end{thebibliography}

\end{document}